\documentclass[final]{aipproc}
\usepackage{graphicx,amsmath,amsfonts,amssymb,mathrsfs}

\layoutstyle{8x11double}

\newcommand{\be}{\begin{eqnarray}}
\newcommand{\ee}{\end{eqnarray}}
\def\nue{{\nu_e}}
\def\anue{{\bar\nu_e}}
\def\numu{{\nu_{\mu}}}

\newcommand{\ma}{\Delta m^2_{31}}

\newcommand{\stch}{\sin^2 2\theta_{13}}

\newcommand{\stcht}{\sin^2 2\theta_{13}{\mbox {(true)}}}

\newcommand{\dcpt}{\delta_{CP}{\mbox {(true)}}}

\begin{document}

\title{Physics with Beta-Beam}



\author{Sanjib Kumar Agarwalla}{
  address={Harish-Chandra Research Institute, Chhatnag Road, Jhusi, Allahabad
  - 211019, India}
}

\author{Sandhya Choubey\footnote{Sandhya Choubey presented this 
plenary talk
at the 9th International Workshop on Neutrino Factories, SuperBeams and 
BetaBeams (NuFact07), Okayama University, Okayama, Japan, August 6-11, 2007.}}{
%  address={<common address for author2 and author3>}
   altaddress={Harish-Chandra Research Institute, Chhatnag Road, Jhusi,
  Allahabad - 211019, India}
}

\author{Amitava Raychaudhuri}{
%  address={<common address for author2 and author3>}
  altaddress={Harish-Chandra Research Institute, Chhatnag Road, Jhusi,
  Allahabad - 211019, India} % additional visiting address
}

\begin{abstract}
A Beta-beam would be  
a high intensity source of pure $\nue$ and/or 
$\anue$ flux with known spectrum, 
ideal for precision measurements.
Myriad of possible set-ups with 
suitable choices of baselines, detectors and the beta-beam
neutrino source with desired energies 
have been put forth in the literature. In this talk we present 
a comparitive 
discussion of the physics reach of a few such 
experimental set-ups.
\end{abstract}

\maketitle

%%%%%%%%%%%%%%%%%%%%%%%%%%%%%%%%%%%%%%%%%%%%
%% MAINMATTER
%%%%%%%%%%%%%%%%%%%%%%%%%%%%%%%%%%%%%%%%%%%%

%%%%%%%%%%%%%%%%%%%%%%
\section{Introduction}
%%%%%%%%%%%%%%%%%%%%%%

Neutrino physics is now poised to move into the precision regime.
A number of high-precision neutrino oscillation experiments
have been contrived to shed light on the third mixing angle $\theta_{13}$,
the sign$\footnote{The neutrino mass hierarchy is
termed ``normal (NH)'' (``inverted (IH)'') if $\ma = m_3^2 - m_1^2$ is
positive (negative).}$ of $\Delta m^2_{31}\equiv m_3^2 - m_1^2$ ($sgn(\ma)$)
and the CP phase ($\delta_{CP}$), key missing 
ingredients of the neutrino mass matrix.  
The $\nu_e \to \nu_\mu$ transition probability 
($P_{e \mu}$) depends 
on all these three parameters and is termed the ``golden channel'' 
\cite{golden,freund} for long baseline accelerator 
based experiments\footnote{The $\nu_e$ survival 
probability can also be used to cleanly measure $\theta_{13}$
and $sgn(\ma)$ \cite{pee}.}. 
In order to exploit this channel, 
we need a pure and intense 
$\nu_e$ (or $\bar\nu_e$) 
beam at the source. The beta-beam 
serves this purpose. In this talk, we will focus on a 
few proposed 
experimental scenarios dealing with beta-beam
and discuss the consensus direction 
for the future.   

%%%%%%%%%%%%%%%%%%%%%%%%%
\section{BETA-BEAM}
%%%%%%%%%%%%%%%%%%%%%%%%%

Zucchelli \cite{zucc} put forward the novel idea of beta-beam
\cite{pee,zucc,lowtohigh,betapemu,cernmemphys,abhijit,betaino,rathin,
subhendu,newcernino},
which is based on the concept of creating a
pure, well-known, intense, collimated beam of $\nu_{e}$ or $\bar\nu_{e}$
through the beta decay of completely ionized radioactive ions. It
will be achieved by producing, collecting, and accelerating these
ions and then storing them in a ring \cite{jacques}. 
Feasibility of this proposal and its physics potential
is being studied in depth \cite{iss},
and will take full 
advantage of the existing 
accelerator complex and CERN and FNAL. 
%The main future challenge lies in
%building an intense proton driver and the hippodrome-shaped
%decay ring which are essential for this programme.
It has been proposed to produce $\nu_e$ beams through the decay of highly
accelerated $^{18}$Ne ions and ${\bar \nu_e}$ from $^6$He
\cite{zucc,jacques}. More recently, $^{8}$B and $^{8}$Li 
\cite{rubbia} with much larger 
end-point energy have been suggested as alternate sources since these 
ions can yield higher energy $\nu_e$  and $\bar\nu_e$ respectively, 
with lower values of the Lorentz boost $\gamma$ 
\cite{rathin,subhendu,doninialter}.  
It may be possible to store radiactive ions producing 
beams with both polarities in
the same ring. 
This will enable running the experiment in the 
$\nue$ and $\anue$ modes simultaneously.
Details 
of the four beta-beam candidate ions
can be found in Table 1 of \cite{betaino}.

%%%%%%%%%%%%%%%%%%%%%%%%%%%%%%%%%%%%%%%%%%%%%%%%%%%%%%%%%%%%%%%%%%%%
\section{The ``GOLDEN CHANNEL'' ($\nu_e$ $\rightarrow$ $\nu_{\mu}$)}
%%%%%%%%%%%%%%%%%%%%%%%%%%%%%%%%%%%%%%%%%%%%%%%%%%%%%%%%%%%%%%%%%%%%

The expression for $P_{e \mu}$ in matter,
upto second order in the small parameters $\alpha \equiv
\Delta m_{21}^2/\Delta m_{31}^2$ and $\theta_{13}$, 
is \cite{golden,freund}:
{\small{
\begin{eqnarray}
P_{e\mu} &\simeq& \sin^2 2\theta_{13} \, \sin^2 \theta_{23}
\frac{\sin^2[(1- \hat{A}){\Delta}]}{(1-\hat{A})^2}\nonumber \\
&+& \alpha  \sin 2\theta_{13} \,  \xi \sin \delta_{CP}
\sin({\Delta})  \frac{\sin(\hat{A}{\Delta})}{\hat{A}}
\frac{\sin[(1-\hat{A}){\Delta}]}{(1-\hat{A})}\nonumber \\
&+& \alpha  \sin 2\theta_{13} \,  \xi \cos \delta_{CP}
\cos({\Delta})  \frac{\sin(\hat{A}{\Delta})}{\hat{A}}
\frac{\sin[(1-\hat{A}){\Delta}]} {(1-\hat{A})}\nonumber \\
&+& \alpha^2 \, \cos^2 \theta_{23}  \sin^2 2\theta_{12}
\frac{\sin^2(\hat{A}{\Delta})}{\hat{A}^2};
\label{eqn:magic}
\end{eqnarray}
}}
where $\Delta \equiv \Delta m_{31}^2 L/(4 E)$, 
$\xi \equiv \sin 2\theta_{12} \, 
\sin 2\theta_{23}$, and $\hat{A} \equiv \pm (2 \sqrt{2}
G_F n_e E)/\Delta m_{31}^2$.  
$G_F$ and $n_e$ are the Fermi coupling constant
and the electron density in matter, respectively. The sign of $\hat{A}$ is
positive (negative) for neutrinos (anti-neutrinos) with NH
and it is opposite for IH. While the 
simultaneous dependence of this oscillation channel on 
$\theta_{13}$, $sgn(\ma)$ and $\delta_{CP}$ allows for 
the simulataneous measurement of all these three quantities, 
it also brings in the problem of ``parameter  
degeneracies'' -- 
the $\theta_{13}$-$\delta_{CP}$ intrinsic degeneracy \cite{intrinsic},
the $sgn(\ma)$ degeneracy \cite{minadeg} and the octant of $\theta_{23}$ 
degeneracy \cite{th23octant} -- leading to an overall eight-fold degeneracy 
in the parameter values \cite{eight}. The degeneracies, unless 
tackled, always reduce the sensitivity of the experiment. 

%%%%%%%%%%%%%%%%%%%%%%%%%%%%%%%%%%%%%%%%%%%
\section{The CERN-INO MAGICAL SET-UP}
%%%%%%%%%%%%%%%%%%%%%%%%%%%%%%%%%%%%%%%%%%%
%%%%%%%%%%%%%%%%%%%%%%%%%%%%%%%%%%%%%%%%%%%%%%%%%%%%%%%%%%%%%%%
\begin{figure}[!t]

\includegraphics[height=.24\textheight]{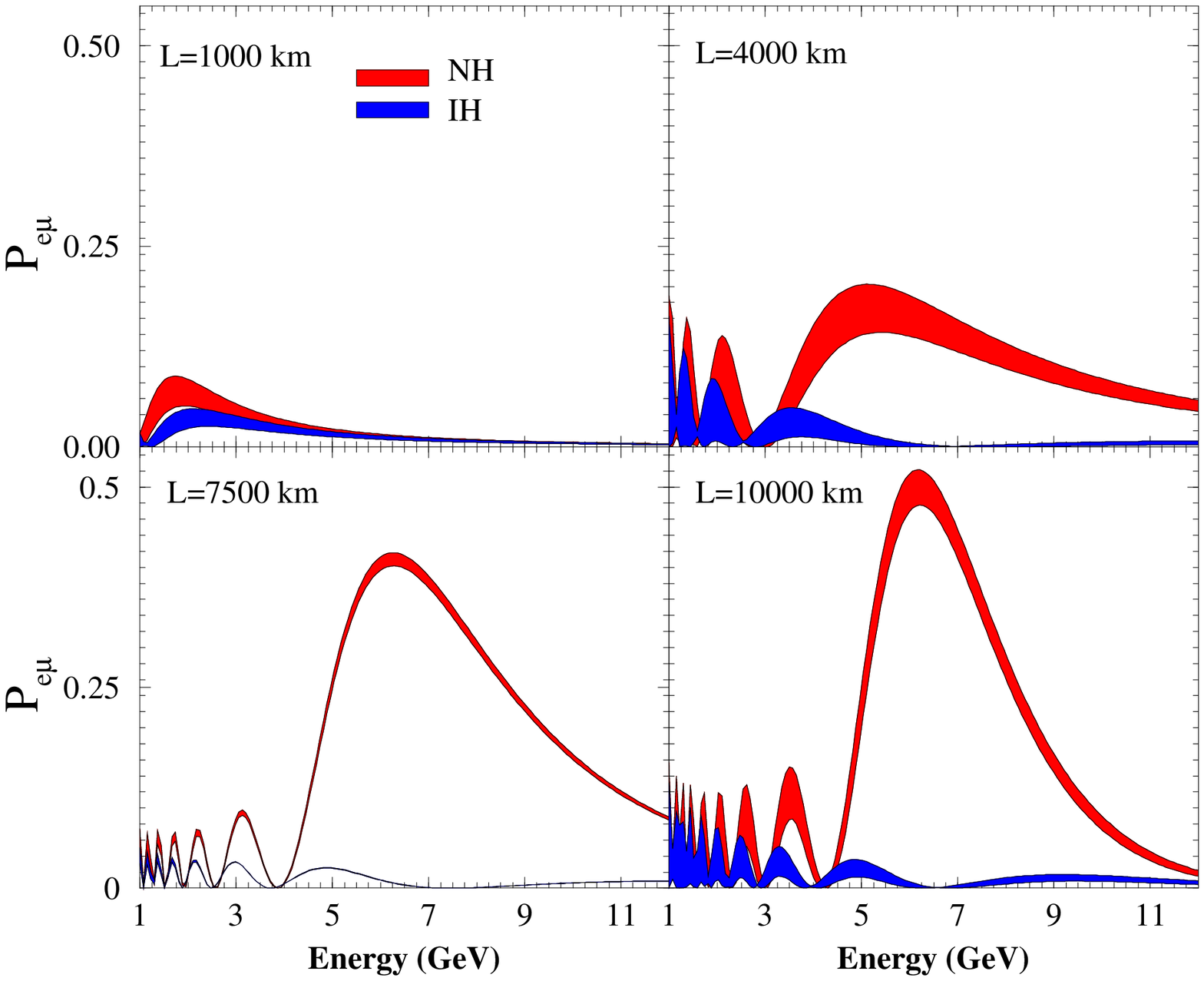}
\hglue 1.5cm
\includegraphics[height=.24\textheight]{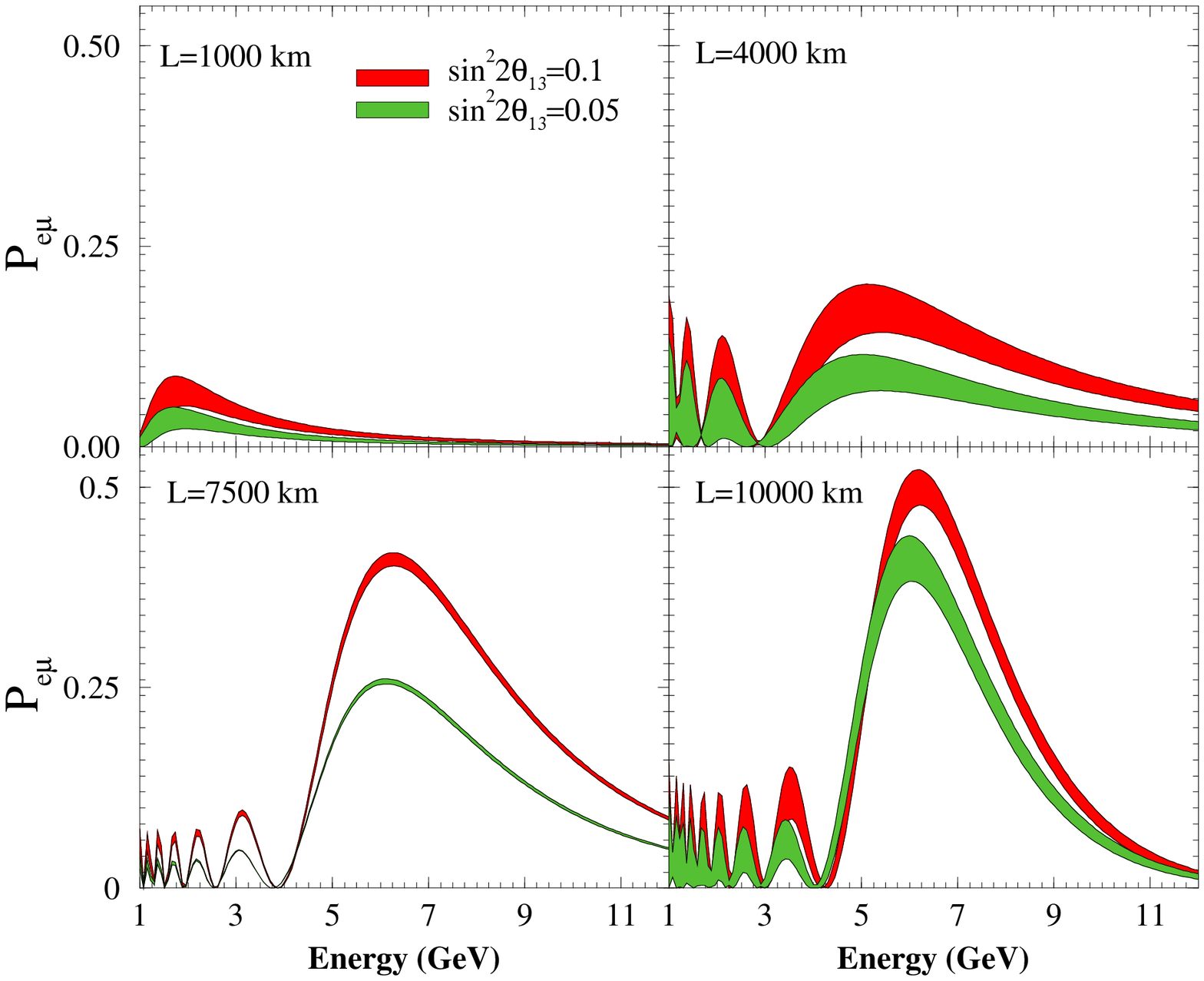}

\caption{\label{fig:prob}
Both the panels show the energy dependence of $P_{e\mu}$
for four baselines where the band reflects the effect of the
unknown $\delta_{CP}$. Left panel clearly depicts the effect of $\delta_{CP}$
in making distinction between normal (NH) $\&$ inverted (IH) hierarchy
with $\stch=0.1$. Right panel reflects the difference in $P_{e\mu}$
for two different values of $\stch$ with NH.}
\end{figure}
%%%%%%%%%%%%%%%%%%%%%%%%%%%%%%%%%%%%%%%%%%%%%%%%%%%%%%%%%%%%%%%%%%%

%%%%%%%%%%%%%%%%%%%%%%%%%%%%%%%%%%%%%%%%%%%%%%%%%%%%%%%%%%%
\begin{figure}[!t]

\includegraphics[height=.24\textheight]{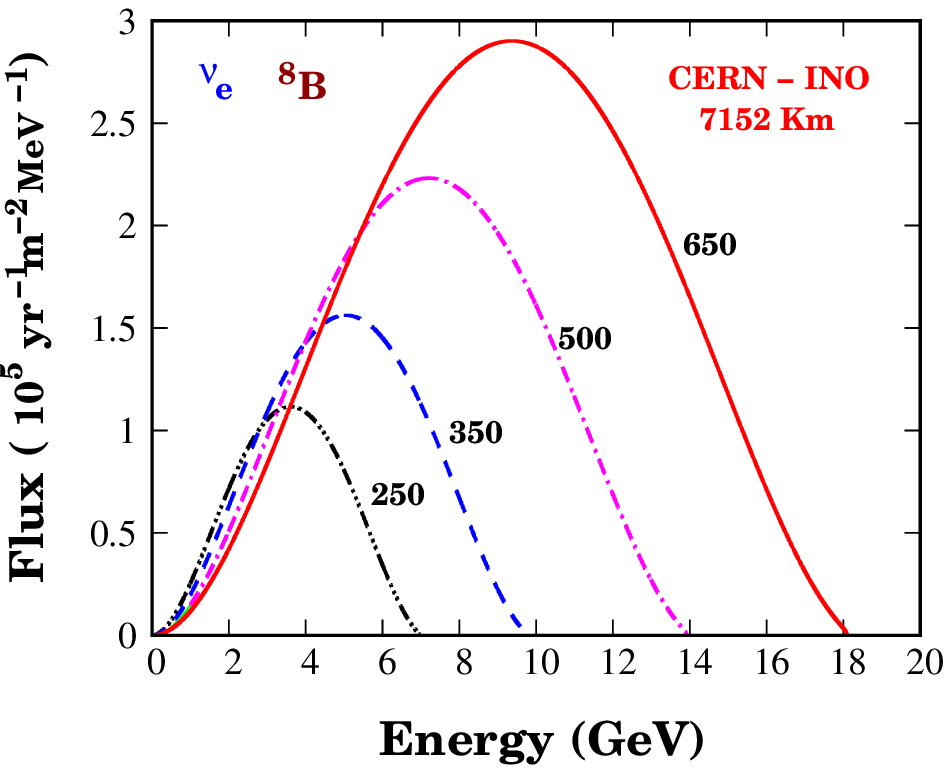}
\hglue 1.0cm
\includegraphics[height=.24\textheight]{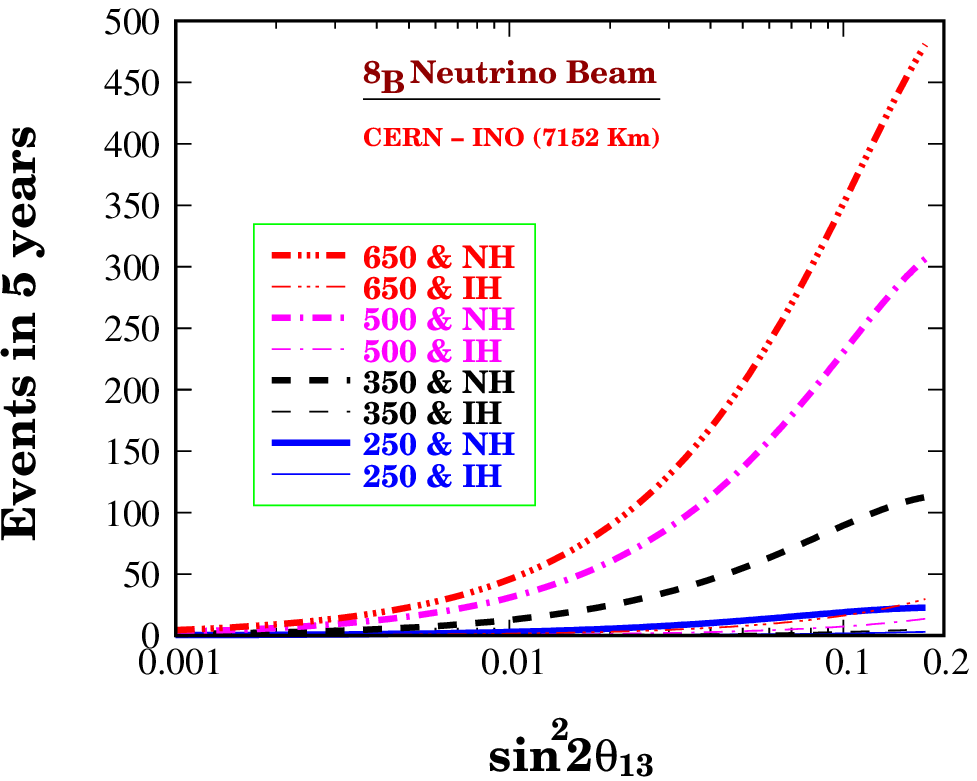}

\caption{\label{fig:flux_rate}
Left panel shows the boosted unoscillated spectrum of neutrinos
from $^8B$ ion which will hit the INO detector, for four different
benchmark values of $\gamma$. The expected number of $\mu^{-}$ events in
5 years running time with 80\% detection efficiency as a function of
$\stch$ are presented in right panel. The value of $\gamma$ and
the hierarchy chosen corresponding to each curve is shown in the
figure legend.}
\end{figure}
%%%%%%%%%%%%%%%%%%%%%%%%%%%%%%%%%%%%%%%%%%%%%%%%%%%%%%%%%%%%%%

Interestingly, when 
$\sin(\hat{A}\Delta)=0$, the last three 
terms in Eq. (\ref{eqn:magic})
drop out and the $\delta_{CP}$ dependence 
disappears from the $P_{e\mu}$ channel. 
The problem of clone solutions due to the 
first two types of degeneracies are therefore evaded.
Since $\hat{A}\Delta = \pm  (2 \sqrt{2} G_F n_e L)/4$
by definition, the first non-trivial solution for 
$\sin(\hat{A}\Delta)=0$ reduces to $\rho L = \sqrt{2}\pi/G_F Y_e$,
where $Y_e$ is the electron fraction inside earth. This
gives $\frac{\rho}{[{\rm g/cc}]}\frac{L}{[km]} \simeq 32725~$,
which for the PREM \cite{prem} density profile of the earth
is satisfied for the ``magic baseline'' \cite{eight,magic,magic2},
$L_{\rm magic} \simeq 7690 ~{\rm km}$.
At this baseline the sensitivity to the 
mass hierarchy and  
$\theta_{13}$ is quite significant \cite{magic}, 
while the sensitivity to $\delta_{CP}$ is absent.

The large baseline also entails 
traversal of neutrinos through 
denser regions of the earth, 
capturing  near-maximal matter contribution 
to the oscillation probability.
%essential for probing $sgn(\ma)$. 
In fact, for this baseline, 
the average earth matter density calculated using the PREM
profile is $\rho_{av}=4.25$ gm/cc, for which the resonance energy
\be
E_{res} &\equiv& {|\ma| \cos 2\theta_{13} \over
2\sqrt{2} G_F N_e}~\\
&=& 7~{\rm GeV}~,
\label{eq:eres}
\ee
for $|\ma|=2.4\times 10^{-3}$ eV$^2$ and $\stch=0.1$.
Of course neutrino oscillation probability for long baseline 
experiments depend on the product of the mixing term and 
the mass squared difference driven oscillatory term inside 
matter. Largest flavor conversions are possible when both these 
terms are large \cite{pee,gandhi}. 
The exact neutrino transition probability $P_{e\mu}$ using the 
PREM density profile is given in Fig. \ref{fig:prob} which has been
taken from \cite{betaino}. 
%The authors allow $\delta_{CP}$ to take on
%all possible values between 0 to $2\pi$ and the resultant probability 
%is shown as a band, with the  thickness of the band reflecting the 
%effect of $\delta_{CP}$ on $P_{e\mu}$.
For neutrinos (antineutrinos),
matter effects for the longer baselines bring a 
significant enhancement of $P_{e\mu}$ for NH (IH), 
while for IH (NH), the probability is almost unaffected.
This feature can be used to determine the neutrino
mass hierarchy (see left panel of Fig. \ref{fig:prob}).
For $L=7500$ km, which is close to the magic baseline, 
the effect of the CP phase is seen to be almost negligible.
This allows a clean measurement of $sgn(\ma)$ and 
$\theta_{13}$
(see right panel of Fig. \ref{fig:prob}),
while for all other cases the impact of $\delta_{CP}$ on $P_{e\mu}$ is
appreciable. 

A large magnetized iron calorimeter (ICAL) is all set to come 
up at the India-based Neutrino Observatory (INO) \cite{ino}. 
ICAL@INO will be a 50 kton detector, 
capable of detecting muons along with their charge, 
with good energy and angular resolution. It might be upgraded 
to 100 kton. 
If a beta-beam facility is built at CERN, ICAL@INO could 
serve as an excellent far detector for observing the oscillated 
$\numu$. The USP of this experimental set-up would be 
the CERN-INO distance, which corresponds to 7152 km, 
tantalizingly close to the magic baseline. This would 
enable an almost degeneracy-free measurement 
of $sgn(\ma)$ and $\theta_{13}$ as discussed above. 
In addition, one could exploit the near-maximal matter effects 
by tuning the beam energy to be close to 6-7 GeV (see Fig. 
\ref{fig:prob}). 

We consider $^{8}$B ($^{8}$Li) 
\cite{rubbia} ion as a possible source for a $\nu_e$ ($\bar{\nu}_e$)
beta-beam and show the expected flux for our experimental 
set-up in the left panel of Fig. \ref{fig:flux_rate}. 
For the Lorentz boost factor $\gamma=250-650$
the $^{8}$B and $^{8}$Li sources have peak energy
around $\sim 4-9$ GeV. We assume 
$2.9\times 10^{18}$ useful decays per year for $^8$Li and
$1.1\times 10^{18}$ for $^8$B, for all values of $\gamma$.
The expected number of events are shown in 
the right panel of Fig. \ref{fig:flux_rate}.
We take a detector energy threshold of 1.5 GeV, 
detection efficiency of 80\% and 
charge identification efficiency of 95\%. 
For 
discussion on our backgrounds and 
details of our statistical analysis
we refer the readers to 
\cite{betaino,newcernino}.

%%%%%%%%%%%%%%%%%%%%%%%%%%%%%%%%%%%%%%%%%%%%%%%%%%%%%%%%%%%%%%
\begin{figure}[!t]

\includegraphics[height=.24\textheight]{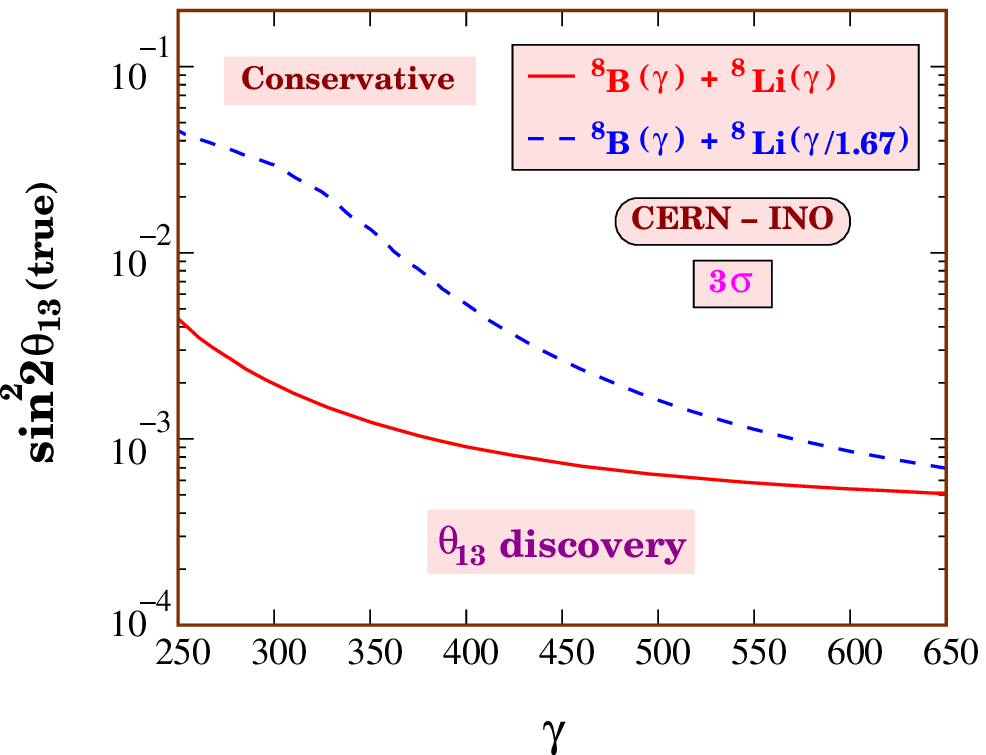}
\hglue 1.0cm
\includegraphics[height=.24\textheight]{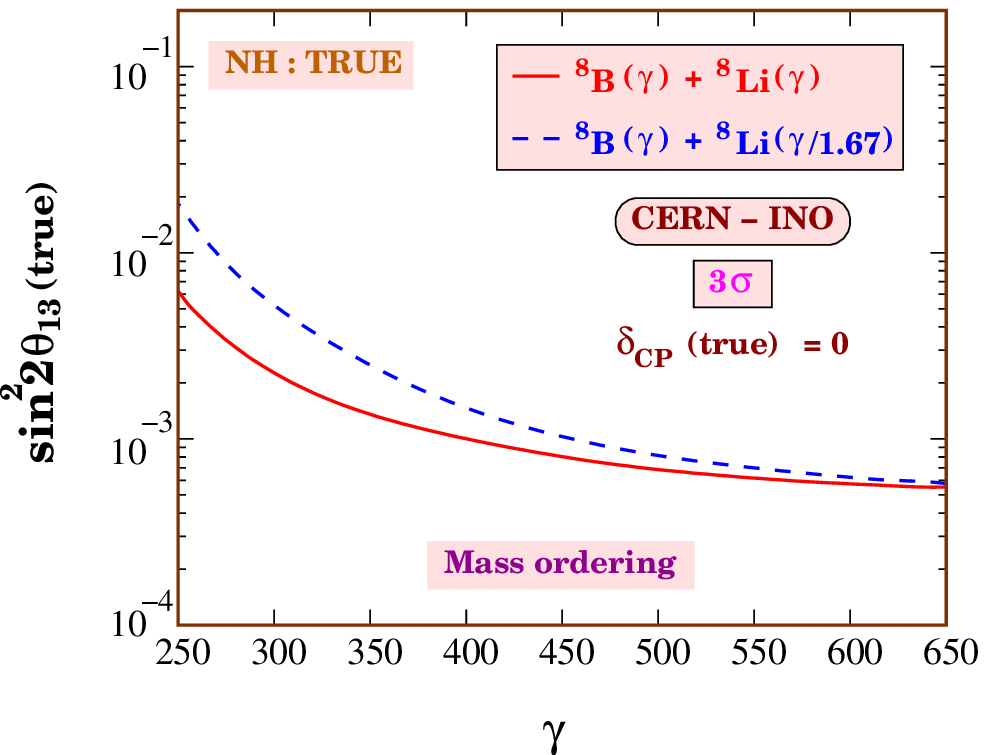}

\caption{\label{fig:senshier}
Left panel shows the $3\sigma$ discovery reach for $\stcht$. Right
panel shows the minimum value of $\stch$(true) for which the wrong 
inverted hierarchy can be ruled out at the $3\sigma$ C.L., as a function of
the Lorentz boost $\gamma$. The red solid lines in both the panels are obtained 
when the $\gamma$ is assumed to be the same for both the neutrino and 
the antineutrino beams. The blue dashed lines
show the corresponding limits when the $\gamma$ for the
$^8$Li is scaled down by a factor of 1.67 with respect to the
$\gamma$ of the neutrino beam, which is plotted in the $x$-axis.
}
\end{figure}
%%%%%%%%%%%%%%%%%%%%%%%%%%%%%%%%%%%%%%%%%%%%%%%%%%%%%%%%%%%%%%%%%%%%

We define the $\stch$ 
sensitivity reach of the CERN-INO beta-beam experiment
as the upper limit on $\stch$ that can be put at the $3\sigma$ C.L.,
in case no signal for $\theta_{13}$ driven oscillations is observed and the
data is consistent with the null hypothesis. 
%The result is shown in the 
%left panel of Fig. \ref{fig:senshier}, as a function of $\gamma$ after
%marginalizing over hierarchy and all other oscillation parameters.   
At $3\sigma$, the CERN-INO $\beta$-beam set-up can constrain
$\sin^22\theta_{13} < {1.14\times 10^{-3}}$ with 
five years of running of the beta-beam in 
both polarities with the same ${{\gamma=650}}$ 
and full spectral information. The $\stcht$ discovery reach 
is defined as the minimum value of $\stcht$ for which we 
can distinguish the signal at the $3\sigma$ C.L. We present 
our results in the left panel of Fig. \ref{fig:senshier}, 
as a function of $\gamma$. The plot presented show
the most conservative numbers which have been obtained 
by considering all values of $\dcpt$ and both hierarchies. 
We refer the reader to \cite{newcernino} for details. 
The hierarchy sensitivity is defined 
as the minimum value of $\stcht$,
for which one can rule out the wrong hierarchy at $3\sigma$ C.L. 
The results are depicted as a function of $\gamma$ in the right
panel of Fig. \ref{fig:senshier}. For NH true, the 
$sgn(\ma)$ reach corresponds to 
$\stcht > {{5.51 \times 10^{-4}}}$, with 
5 years energy binned data of both polarities and $\gamma=650$.
Here we had assumed $\dcpt=0$. However, as discussed before, 
the effect of $\delta_{CP}$ is minimal close to the magic 
baseline and hence we expect this sensitivity to be almost 
independent of $\dcpt$ (see \cite{newcernino} for details).

%%%%%%%%%%%%%%%%%%%%%%%%%%%%%%%%%%%%%%%%%%%%%%%%%%%%%%%%%%%%%
\section{The CERN-MEMPHYS PROJECT}
%%%%%%%%%%%%%%%%%%%%%%%%%%%%%%%%%%%%%%%%%%%%%%%%%%%%%%%%%%%

%%%%%%%%%%%%%%%%%%%%%%%%%%%%%%%%%%%%%%%%%%%%%%%%%%%%%%%%
\begin{figure}
\includegraphics[width=\columnwidth]{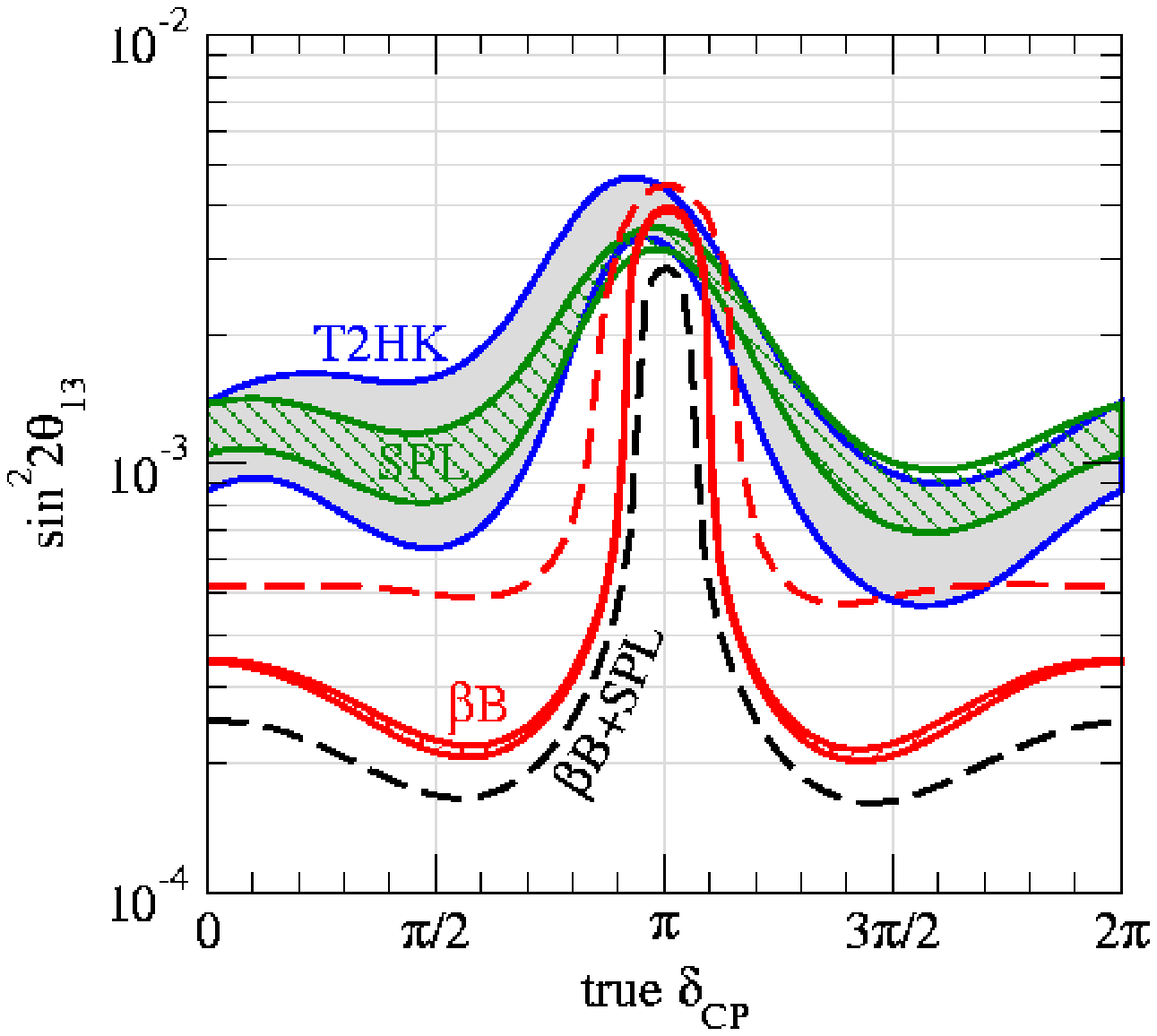}
\caption{\label{fig:cernmemphys}
$3\sigma$ discovery reach for $\stcht$ for $\beta$-beam,
Super Beam and T2HK (phase~II of the T2K experiment) as a function of
$\dcpt$. The running time is ($5\nu + 5\bar\nu$)~year for $\beta$-beam
with twice the standard luminosity and ($2\nu + 8\bar\nu$)~years for
the Super Beams (4 MW).
}
\end{figure}
%%%%%%%%%%%%%%%%%%%%%%%%%%%%%%%%%%%%%%%%%%%%%%%%%%%%%%%%%%

The CERN-MEMPHYS proposal comprises of sending a low 
gamma beta-beam from CERN to the envisaged MEMPHYS, 
which would be a 440 kton fiducial mass 
water detector
located in Fr\'ejus, at a distance of 130 km from CERN.
The major advantage of this set-up is that one needs
very reasonable values of the Lorentz Boost $\gamma=100$ and  
$^{18}$Ne and $^6$He ions for producing the beta-beam. 
The current accelerator capabilities at CERN are expected to be 
enough for producing a beta-beam with $\gamma=100$ without 
requiring any upgrades and affecting the running of LHC.  
The band between the red solid lines in 
Fig. \ref{fig:cernmemphys} show the $3\sigma$ 
``discovery reach'' 
for $\stcht$ using the combined 5 years run in  
$\nue$ and $\anue$ polarities. The band corresponds to 
changing the systematic errors from 2\% to 5\%. 
The $3\sigma$ 
$\stcht$ discovery reach is 
defined as the minimum value of $\stcht$ which 
could produce a $3\sigma$ unambiguous signal at the detector.
The strongest point of this experiment is its 
tremendous sensitivity to CP violation. Maximal CP violation 
can be observed at the $3\sigma$ C.L. if 
$\stcht > 2\times 10^{-4}$.   
Another major advantage of this set-up is that if the SPL 
is built at CERN, then it could serve as a superbeam experiment 
as well. In that case, one could run could combine simultaneous
5 years of running of $\nue$ beta-beam with 5 years of running of 
the SPL superbeam, without having to run the experiment 
in the $\anue$ mode.

%%%%%%%%%%%%%%%%%%%%%%%%%%%%%%%%%%%%%%%%%%%%%%%%%%%%%%%%%%%%%
\section{Comparing Different Set-ups}
%%%%%%%%%%%%%%%%%%%%%%%%%%%%%%%%%%%%%%%%%%%%%%%%%%%%%%%%%%%
%\begin{table}
%\caption{
%Comparison between the different experimental set-ups. 
%See the text for details.
%\end{table}
%% %%%%%%%%%%%%%%%%%%%%%%%%%%%%%%%%%%%%%%%%%%%%%%%%%%%%%%%%%%%%%
 \begin{table}
 \begin{tabular}{lrrrr}\hline
 Set-up&1&2&3\\ \hline
 Detector type&WC&TASD&TASD\\ \hline
 $m\,[\mathrm{kt}]$&500&50&50 \\
 $\gamma$&200&500&1000\\
 $L\,[\mathrm{km}]$&520&650&1000\\ \hline
 $\nu$ signal&1983&2807&7416\\
 $\nu$ background&105&31&95\\ \hline
 \end{tabular}
 \caption{\label{tab:events}
 The number of signal/background events for different combinations
 of the chosen detector type and values of $\gamma$.
WC stands for Water Cherenkov, while TASD means a Totally Active 
Scintillator Detector.}
 \end{table}
%% %%%%%%%%%%%%%%%%%%%%%%%%%%%%%%%%%%%%%%%%%%%%%%%%%%%%%%%%%%%%%%

% %%%%%%%%%%%%%%%%%%%%%%%%%%%%%%%%%%%%%%%%%%%%%%%%%%%%%%%%%%%%%%
 \begin{figure}[t]
 \includegraphics[width=\columnwidth]{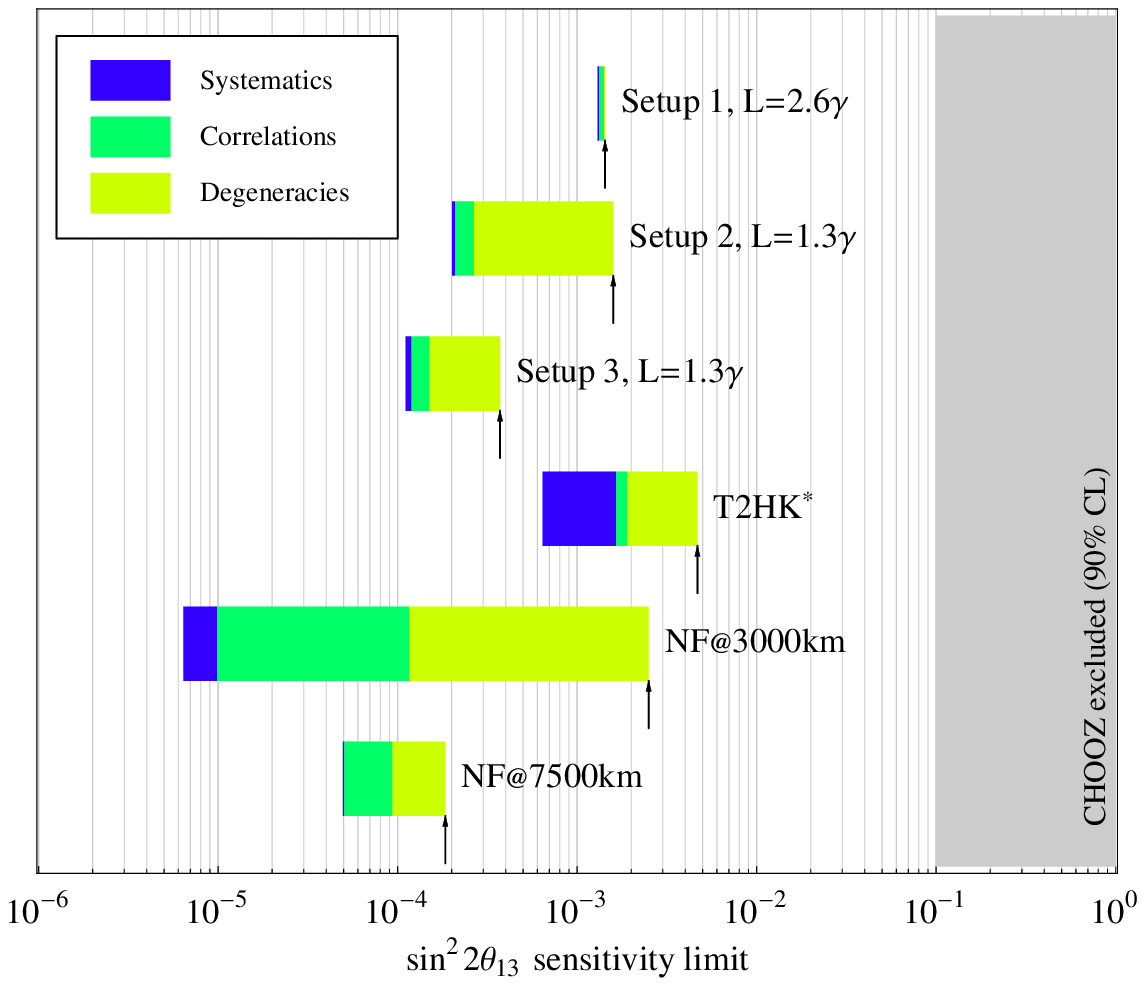}
 \caption{\label{fig:cpf}
 The $\stch$ sensitivity limits for the
 different setups and other representatives. Here $n=0$ (decays per year
 fixed) and the $3 \sigma$ confidence level are chosen. The final sensitivity
 limits are obtained as the right edges of the bars after successively
 switching on systematics, correlations, and degeneracies.
 }
 \end{figure}
% %%%%%%%%%%%%%%%%%%%%%%%%%%%%%%%%%%%%%%%%%%%%%%%%%%%%%%%%%%%%%%%%

The authors of \cite{lowtohigh} studied the physics potential of
beta-beams, using $^{18}$Ne and $^6$He as the source ions and 
allowing for different values of $\gamma$ and $L$. 
Table \ref{tab:events} describes the details of the 
three illustrative set-ups analyzed in details in 
\cite{lowtohigh}. Fig.  \ref{fig:cpf} shows the 
$\stch$ sensitivity reach of these three set-ups and 
compares them with the corresponding potential of 
that expected from two standard neutrino factory set-ups.
We note that the sensitivity of the CERN-INO beta-beam 
experiment is better than that
quoted for the set-up 2 of Table \ref{tab:events}. 
The set-up 3 is better, but it needs $\gamma$ = 1000.
While none of these three set-ups are competitive with 
the neutrino factory at magic baseline or the CERN-INO 
beta-beam set-up as far as the 
hierarchy sensitivity is concerned, 
the CP sensitivity of the three set-ups is extremely good.
For CP studies the performance 
of beta-beam is comparable with neutrino factory at 
$L=3000-4000$ km. 

\begin{figure}[h]
\hglue -0.7cm
\includegraphics[width=\columnwidth]{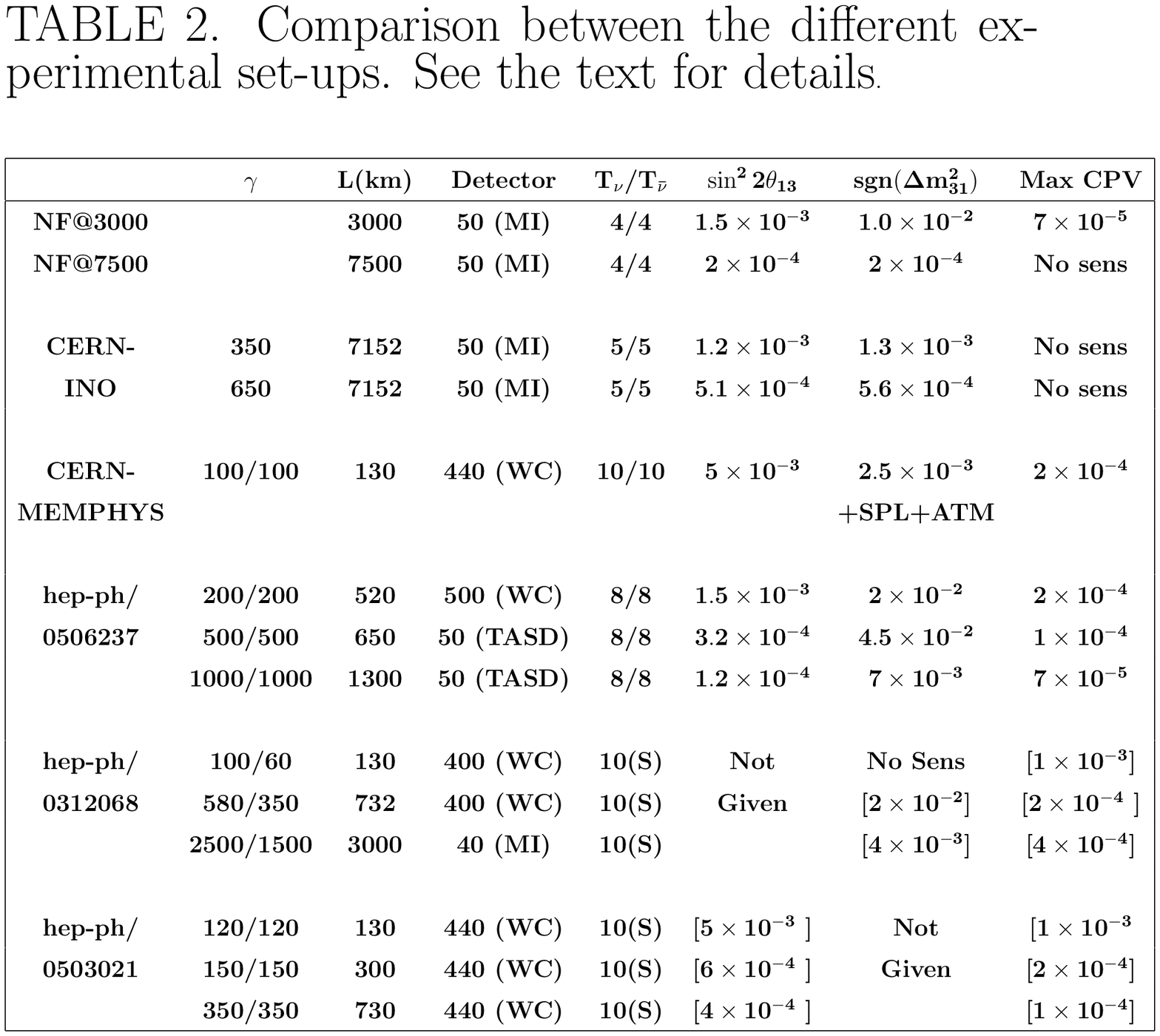}
\end{figure}

In Table 2 
we present a quantitative comparison of the potential 
of the different set-ups. The first two rows 
of the table shows the sensitivity reach of the 
the neutrino factory experiments at 3000 km and 7500 km
respectively. The third and fourth rows show the physics 
reach of the CERN-INO and CERN-MEMPHYS beta-beam proposal. 
The remaining entries have been taken from various 
papers on beta-beam and their arXiv numbers are mentioned 
in the first column of the Table. 
The second column shows the $\gamma$ value considered, 
the third column gives the $L$ taken, 
fourth column the type of detector 
considered\footnote{The detector type (MI) stands for magnetized iron.}, 
while 
the fifth column shows the time of running of the 
experiment in the neutrino ($T_{\nue}$) and 
antienutrino ($T_{\anue}$) modes. 
The cases shown as 10(S) correspond to 
{\it simultaneous} running of the $\nue$ and $\anue$ 
beams for a period of 10 years, with the $\gamma$
corresponding to the $\anue$ beam suppressed by a factor of 
1.67. The last three 
columns show the (approximate) $3\sigma$ 
$\theta_{13}$ discovery (or sensitivity reach), 
the hierarchy sensitivity and CP sensitivity respectively.
The entries in square brackets correspond to 99\% C.L. 
sensitivity. 
The results correspond to assumed true normal hierarchy. 
Since the $\theta_{13}$ and hierarchy reach of the experiment 
in general depends on $\dcpt$, we give the most conservative 
value. Note that for the CERN-MEMPHYS project the hierarchy 
sensitivity comes mainly from adding the atmospheric 
neutrino data in the megaton MEMPHYS detector.

%%%%%%%%%%%%%%%%%%%%%%%%%%%%%%%%%%%%%%%%%
\section{Conclusions}
%%%%%%%%%%%%%%%%%%%%%%%%%%%%%%%%%%%%%%%%%

In this talk, we discussed the expected physics reach of
selected experimental set-ups using a beat-beam. 
Beta-beams are seen to have extremely good physics reach 
which are comparable to those expected in neutrino factories.

%%%%%%%%%%%%%%%%%%%%%%%%%%%%%%%%%%%%%%%%%%%%%%%%%%%%%%%%%%%%%%%%%%%%

%%%%%%%%%%%%%%%%%%%%%%%%%%%%%%%%%%%%%%%%%%%%%%%%%%%%%%%%%%%%%%%%%%%%

\end{document}